\documentclass[acus]{JAC2000}
\usepackage[final]{graphics}
\usepackage{citesort}

\newcommand{\CIPTGGVL} {\ensuremath{-0.003}}
\newcommand{\CIPTGGEM} {\ensuremath{ 0.011}}
\newcommand{\CIPTSSVL} {\ensuremath{ 0.0012}}
\newcommand{\CIPTSSEM} {\ensuremath{ 0.0047}}
\newcommand{\CIPTSEVL} {\ensuremath{-0.0010}}
\newcommand{\CIPTSEEM} {\ensuremath{ 0.0029}}
\newcommand{\CIPTCSMT} {\ensuremath{0.16}}
\newcommand{\CIPTcCHSQ}{\ensuremath{0.63}}
\newcommand{\MEANcGGVL}{\ensuremath{ 0.001}}
\newcommand{\MEANASVLMT}{\ensuremath{ 0.348}}
\newcommand{\MEANASXMMT}{\ensuremath{ 0.009}}
\newcommand{\MEANcVSVL}{\ensuremath{ 0.0256}}
\newcommand{\MEANcVEVL}{\ensuremath{-0.0080}}
\newcommand{\MEANcASVL}{\ensuremath{-0.0197}}
\newcommand{\MEANcAEVL}{\ensuremath{ 0.0041}}

\newcommand{\alphas}    {\ensuremath{\alpha_{\rm s}}}
\newcommand{\Pgne}      {\ensuremath{\mathrm{\nu_e}}}
\newcommand{\Pgngt}     {\ensuremath{\mathrm{\nu_\tau}}}

\newcommand{\Ph}        {\ensuremath{\mathrm{h}}}
\newcommand{\Pe}        {\ensuremath{\mathrm{e}}}
\newcommand{\Pep}       {\ensuremath{\mathrm{e^+}}}
\newcommand{\Pem}       {\ensuremath{\mathrm{e^-}}}

\title{\flushright{M09}\\[-2mm] \centering $\alpha_{\rm\bf s}$ AT LOW
  $Q^2$ FROM $\Pep\Pem$ AND $\tau$ DATA\thanks{Work supported by
    Department of Energy contract DE--AC03--76SF00515.}}

\author{S. Menke, Stanford Linear Accelerator Center,
    2575 Sand Hill Road, Menlo Park, CA 94205, USA}
\begin{document}

\maketitle

\begin{abstract} It has been shown in recent analyses by 
  ALEPH~\cite{art:Andreas98} and OPAL~\cite{art:Sven} that precision
  QCD tests are possible with hadronic $\tau$ decays by comparing
  spectral moments of the hadronic decay ratio of the $\tau$ with QCD
  calculations.  In principle $\Pep\Pem$ data can be used in a similar
  manner by evaluating spectral moments of $R$.  The current
  $\Pep\Pem$ data is compared with the OPAL $\tau$ data and a
  prediction is made on the achievable accuracy of QCD tests with the
  projected precision of PEP-N~\cite{art:PEP-N}.
\end{abstract}


\section{INTRODUCTION}
The $\tau$ lepton is the only lepton heavy enough to decay into
hadrons. The observed spectra of the non-strange hadrons in $s$, where
$\sqrt{s}$ is the mass of the final state hadronic system, give the
non-strange spectral functions.  Decays with an even number of pions
in the final state belong to the vector current while decays with an
odd number of pions belong to the axial-vector current. A comparison
of weighted integrals over the spectral functions of the vector and
axial-vector current with QCD predictions can give fundamental
parameters of the
theory\cite{art:Braaten88,art:Braaten89,art:Narison88,art:Braaten,art:Pich92},
including the strong coupling constant \alphas. Different integrals
(moments) are used to measure power corrections of non-perturbative
origin and the strong coupling simultaneously, thus reducing the
theoretical uncertainties on \alphas{\ }connected with the
non-perturbative terms.

The QCD tests can also be performed for energy scales smaller than
$m_\tau$, if the integrals over the spectral functions are performed
with an upper integration limit $m_{\tau'}^2 = s_0\le m_\tau^2$ and
replacing $m_\tau^2$ by $m_{\tau'}^2$ in the
integrals~\cite{art:Andreas98} thus creating the spectral moments of a
hypothetical $\tau'$ lepton with a mass below $m_\tau$.

In the same manner $\Pep\Pem$ annihilation into hadrons can be used to
extract the spectral function of the vector-current and compare with
theory by means of its spectral moments.


\section{THEORY}
%


\subsection{Hadronic $\tau$ decays}
The vector $v(s)$ and axial-vector $a(s)$ spectral functions (the
absorptive parts of the vacuum polarization correlators) are given by
the spectra in $s$ of the final state hadrons, normalized to the
branching ratios $B$ and corrected for the phase-space:
\begin{eqnarray}
  v/a(s)\! & = & \!\!\frac{1}{2\pi} {\rm Im}\Pi_{\rm V/A}(s) \nonumber\\
           & = & \! \!\frac{\displaystyle
    m_\tau^2 \!\sum_{\Ph_{\rm V/A}} 
    \frac{\displaystyle B_{\tau\rightarrow \Ph_{\rm V/A} \Pgngt}}{\displaystyle
      B_{\tau\rightarrow \Pe \Pgne \Pgngt}}
    \frac{\displaystyle w_{\rm V/A}}{\displaystyle N_{\rm V/A}}
    \frac{\displaystyle {\rm d} N_{\rm V/A}}{\displaystyle {\rm d} s}}
    {\displaystyle 6 S_{\rm EW} |V_{\rm ud}|^2 
    \left(1\! -\! \frac{s}{m_\tau^2}\right)^2
    \left(1\! +\! 2\frac{s}{m_\tau^2}\right)
    }, 
  \label{eq:spectral} 
\end{eqnarray}
where the sum is performed over non-strange hadronic final states
$h_{\rm V/A}$ with angular momentum $J\!=\!1$. $N_{\rm V/A}$ is the
number of taus that decay into the hadron $\Ph_{\rm V/A}$ plus
neutrino, $w_{\rm V/A}$ denotes the appropriate weight of the hadronic
mode to the vector or axial-vector current, $S_{\rm EW} = 1.0194$ is
an electroweak correction term~\cite{art:Marciano} and $|V_{\rm ud}|^2
= 0.9477 \pm 0.0016$ is the squared CKM weak mixing matrix
element~\cite{art:PDG00}.

Within the framework of QCD weighted integrals over the spectral
functions (so called moments) have been
calculated~\cite{art:Diberder}:
\begin{eqnarray}
  \lefteqn{R_{\tau,{\rm V/A}}^{kl}\! = 6 S_{\rm EW} |V_{\rm ud}|^2\!\!
    \int\limits_0^{m_\tau^2}\! 
    \frac{{\rm d} s}{m_\tau^2}
    \left( 1\! -\! \frac{s}{m_\tau^2}\right)^{2+k}\!
    \left(\frac{s}{m_\tau^2}\right)^l} \nonumber\\  
  & & \phantom{R_\tau =} \times
  \left[ \left(1\!+\!2\frac{s}{m_\tau^2}\right) v/a(s) + v^0/a^0(s)
  \right]\label{eq:rkl}, 
\end{eqnarray}
where the scalar spectral function $v^0(s)$ vanishes, since no scalar
particle has been observed in $\tau$ decays, while the pseudo-scalar
spectral function $a^0(s)$ is given by the pion pole, assuming that
the pion is the only pseudo-scalar final-state in non-strange $\tau$
decays:
\begin{equation}
  a^0(s) = 
  \frac{\displaystyle m_\tau^2 
    \frac{\displaystyle B_{\tau \rightarrow \pi \Pgngt}}{
      \displaystyle B_{\tau \rightarrow \Pe \Pgne \Pgngt}}
    \delta(s-m_\pi^2)}
  {\displaystyle 6 S_{\rm EW} |V_{\rm ud}|^2
    \left(1\!-\!\frac{s}{m_\tau^2}\right)^{2}}.
  \label{eq:pionpole}
\end{equation}

The moments are used to compare the experiment with theory. In what
follows, ten moments for $kl = 00,10,11,12,13$ for $\rm V$ and $\rm A$
are used.  The first moments~$R_{\tau,{\rm V/A}}^{00}$ are the total
normalized decay rates of the $\tau$ into vector and axial-vector
mesons. In the na{\"\i}ve parton model these two rates are identical and
add up to the number of colors. Since only non-strange currents are
considered in this work the na{\"\i}ve expectation has to be multiplied by
$|V_{\rm ud}|^2$.  Including the perturbative and non-perturbative
contributions, equation (\ref{eq:rkl}) is usually written as
\cite{art:Diberder}:
\begin{equation}
  R_{\tau,{\rm V/A}}^{kl}\! = 
  \frac{3}{2} S_{\rm EW} |V_{\rm ud}|^2 
  \biggl( 1 +
  \delta_{\rm pert}^{kl} + \!\!\!\!\!\!\sum_{D = 2,4,6,8}\!\!\!\!\!\!
  \delta_{\rm V/A}^{D,kl} \biggr),        
  \label{eq:rklQCD}      
\end{equation}
where $\delta_{\rm pert}^{kl}$ are perturbative QCD corrections
($\approx 20\,\%$ for $kl\!=\!00$) and the $\delta_{\rm V/A}^{D,kl}$
terms are the so-called power corrections ($\approx 1\,\%$ for
$kl\!=\!00$).

The perturbative term $\delta_{\rm pert}^{kl}$ is known to third order
in \alphas{\ }and partly known to fourth order in \alphas. For $kl=00$
the Contour Improved Perturbation Theory (CIPT) result
is~\cite{art:Diberder}:
\begin{eqnarray}
  \delta_{\rm pert}^{00} = \sum_{n=1}^4 \frac{K_n}{2\pi i} \!\!\!
  \oint\limits_{|s|=m_\tau^2} \!\!\!\frac{{\rm d} s}{s}
  \left(1-2\frac{s}{m_\tau^2}+2\frac{s^3}{m_\tau^6}\right.\nonumber\\
  \left.-\frac{s^4}{m_\tau^8}\right)
  \left(\frac{\alphas(-s)}{\pi}\right)^n,\label{eq:CIPT}
\end{eqnarray}
where the $K_1=1, K_2=1.63982, K_3=6.37101$ are
known~\cite{art:Chetyrkin,art:Dine,art:Celmaster,art:Gorishnii,art:Surguladze}
and $K_4 = 25\pm50$ has been
estimated~\cite{art:Diberder,art:Pich97,art:Kataev}.  The
Taylor-expansion in $\alphas(m_\tau^2)$ of the CIPT result
is~\cite{art:Braaten,art:Diberder}:
\begin{equation}
  \label{eq:FOPT}
  \delta_{\rm pert}^{00} =  a_{\rm s}
  +  5.2023      a^2_{\rm s}
  + 26.366      a^3_{\rm s}   
  + (78.003 + K_4) a^4_{\rm s},
\end{equation}
with $a_{\rm s} = \alphas(m_\tau^2)/\pi$. It is referred to as Fixed
Order Perturbation Theory (FOPT).  The third method considered in this
paper resums the leading term of the $\beta$-function to all orders in
\alphas{\ }by inserting so-called Renormalon Chains
(RCPT)~\cite{art:Ball,art:Neubert,art:Maxwell95,art:Maxwell96}.  The
fixed-order corrected version (up to the third order in \alphas)
quoted in the lower portion of table 6 in reference~\cite{art:Neubert}
is used in the fit.

The power corrections $\delta_{\rm V/A}^{D,kl}$ in the framework of
the Operator Product Expansion (OPE) \cite{art:Wilson} are
proportional to $m_\tau^{-D}$.  The dimension $D=2$ terms are
mass-corrections of perturbative
origin~\cite{art:Braaten,art:Diberder} and are small for non-strange
$\tau$ decays.  Corrections of higher dimension are of
non-perturbative origin, absorbing the long-distance dynamics into
vacuum matrix elements~\cite{art:SVZ1,art:SVZ2,art:SVZ3,art:Braaten}.
In contrast to the perturbative part the power corrections differ for
the vector and the axial-vector currents.

If one neglects the small $s$-dependence of the power corrections, the
$\delta_{\rm V/A}^{D,kl}$ terms can be expressed for all $kl$ values
by a product of the power correction for $kl = 00$ and a simple
integral over the $kl$-dependent weight-functions~\cite{art:Diberder}.
This approach is used for the dimension $D=6$ and $D=8$ terms, taking
$\delta_{\rm V/A}^{6/8,00}$ as free parameters.  For the dimension
$D=2$ and $D=4$ terms the full $s$-dependence is taken into account
for the theoretical description of the moments \cite{art:Diberder}.
The least precisely known $D=4$ parameter, the gluon condensate which
is known only to $50\,\%$ \cite{art:Braaten} is also taken as a free
parameter in the fit, while the $D=2$ term is calculated from the
quark masses and the strong coupling. The quark-masses and
-condensates needed to complete the $D=2,4$ terms are taken
from~\cite{art:Braaten}.


\subsection{$\Pep\Pem$ annihilation into hadrons}
The ratio $R_{\Pep\Pem}$ is defined as:
\begin{eqnarray}
  R_{\Pep\Pem}(s) \!\!& = &\!\! \frac{3s}{4\pi\alpha^2}
  \sigma_{\Pep\Pem\rightarrow\textrm{hadrons}}(s)\nonumber\\
  \!\!&=&\!\! 12\pi\,{\rm Im}\Pi^\gamma(s) 
  = 6 v^\gamma(s),
  \label{eq:R}
\end{eqnarray}
with $v^\gamma(s) \approx\sum_{f={\rm u,d,s}}Q_f^2 v(s)$ being the
vector spectral function with (except for the isoscalar contributions
and the charge dependent factor) similar properties as $v(s)$
in~(\ref{eq:spectral}) and~(\ref{eq:rkl}) and $Q_f$ denoting the
charge of the quark flavor $f$.

In massless perturbative QCD $\sum_f Q_f^2 v(s)$ and $v^\gamma(s)$ are
identical and most conveniently expressed in form of the
Adler-function~\cite{art:Adler}:
\begin{eqnarray}
  D_{\rm P}(-s) &=& -4\pi^2s\frac{{\rm d}\Pi^\gamma(-s)}{{\rm d}s}\nonumber\\ 
  &=& \sum_{f={\rm u,d,s}}\!\!\!Q_f^2\left(1+\sum_{n=1}^4 
    K_n\frac{\alphas^n(s)}{\pi^n}\right)  \label{eq:Adler},
\end{eqnarray}
with (for three flavors) the same $K_n$ as in~(\ref{eq:CIPT}).

The mass corrections to the Adler-function are given
by~\cite{art:Chetyrkin90}:
\begin{eqnarray}
  D_{\rm mass}(-s) & = & \!\!-\!\!\sum_{f={\rm u,d,s}}\!\!\!Q_f^2
  \frac{m_f^2(s)}{s}\Biggl[6+28\frac{\alphas(s)}{\pi}\nonumber\\
  & & \hspace{-1.5cm}+\!\Biggl(\!259.666
  - 2.25\!\!\!\!\!\sum_{f'={\rm u,d,s}}\!\!\!\!\!
  \frac{m_{f'}^2(s)}{m_f^2(s)}\Biggr)
  \frac{\alphas^2(s)}{\pi^2}\Biggr]  \label{eq:Dmass},
\end{eqnarray}
where the running quark masses $m_f(s)$ are calculated from
scale-invariant mass parameters and an evolution equation which is
know to four loops~\cite{art:Vermaseren97}.

Finally the non-perturbative part of the Adler function
reads~\cite{art:Braaten}:
\begin{eqnarray}
  D_{\rm NP}(-s)\!\!\! & = & \!\!\!8\pi^2\!\!\!\!\sum_{f={\rm u,d,s}}\!\!\!Q_f^2
  \left[\frac{1}{12}
    \left(1-\frac{11}{18}\frac{\alphas(s)}{\pi}\right)\!
    \frac{\langle\frac{\alphas}{\pi}GG\rangle}{s^2}\right.\nonumber\\
  & & + \left(2+\frac{2}{3}\frac{\alphas(s)}{\pi}\right)
  \frac{\langle m_f \bar\psi_f\psi_f\rangle}{s^2}\nonumber\\
  & & + \frac{4}{27}\frac{\alphas(s)}{\pi}\!\!\sum_{f'={\rm u,d,s}}\!\!\!
  \frac{\langle m_{f'} \bar\psi_{f'}\psi_{f'}\rangle}{s^2}\nonumber\\
  & & + \left.\frac{3}{2}\frac{\langle{\cal O}_6\rangle}{s^3} 
    + 2\frac{\langle{\cal O}_8\rangle}{s^4}\right]\label{eq:DNP},
\end{eqnarray}
with the dimension 4 contributions from the gluon condensate
$\langle\alphas/\pi GG\rangle$ and the quark condensates $\langle m_f
\bar\psi_f\psi_f\rangle$ and the dimension 6 and 8 operators
$\langle{\cal O}_6\rangle$ and $\langle{\cal O}_8\rangle$.

The total Adler function is simply the sum of the perturbative
part~(\ref{eq:Adler}), the mass corrections~(\ref{eq:Dmass}) and the
non-perturbative part~(\ref{eq:DNP}):
\begin{equation}
  \label{eq:Dtot}
  D(-s) = D_{\rm P}(-s) + D_{\rm mass}(-s) + D_{\rm NP}(-s).
\end{equation}

Moments of similar form to~(\ref{eq:rkl}) can be defined for
$\Pep\Pem$ data~\cite{art:Davier}:
\begin{equation}
  \label{eq:repemkl}
  R_{\Pep\Pem}^{kl}(s_0) = \!\!\!\int\limits_{4m_\pi^2}^{s_0}\!\!
  \frac{{\rm d}s}{s_0}
  \left(1-\frac{s}{s_0}\right)^k\left(\frac{s}{s_0}\right)^l
  R_{\Pep\Pem}(s).
\end{equation}
In order to guarantee the validity of the OPE the endpoint in the
integrals should be suppressed by at least two powers of $s$ and the
moments chosen here are therefore $kl=20,30,31,32,33$. As in $\tau$
decays the moments can be rewritten as integrals around the circle
$|s|=s_0$ in the complex $s$-plane:
\begin{equation}
  \label{eq:rklcirc}
  R_{\Pep\Pem}^{kl}(s_0) = 6\pi i\!\!\!\oint\limits_{|s|=s_0}\!\!
  \frac{{\rm d}s}{s_0}\!
  \left(1-\frac{s}{s_0}\right)^k\!\!\left(\frac{s}{s_0}\right)^l\!\!
  \Pi^\gamma(s),
\end{equation}
and further written as integrals over the Adler function after
integrating by parts:
\begin{eqnarray}
  R_{\Pep\Pem}^{kl}(s_0) & = &\frac{3}{2\pi i}\!\!\!\oint\limits_{|s|=s_0}\!\!\!
  \frac{{\rm d}s}{s}\!
  \sum_{m=0}^k\!\frac{(-1)^m{k\choose m}}{m+l+1}\nonumber\\
  & & \times\left[1-\left(\frac{s}{s_0}\right)^{m+l+1}\right]D(s).  
  \label{eq:rkld}
\end{eqnarray}

In the QCD fits the dimension $D=6$ and $D=8$ terms are fitted using
$\langle{\cal O}_{6,8}\rangle$ as free parameters.  Note that this
differs from the $\tau$ data fits where the contributions to the
moment $kl=00$ are used as free parameters.  For the dimension $D=2$
and $D=4$ terms the full $s$-dependence is taken into account for the
theoretical description of the moments as described above.  In
contrast to the non-strange $\tau$ data where the least precisely
known $D=4$ parameter is the gluon condensate the largest uncertainty
for the $\Pep\Pem$ data comes from the strange quark mass $m_{\rm
  s}(1\,{\rm GeV})$ which is used as a free parameter for the $D=2$
and $D=4$ terms. The other quark-masses and -condensates needed to
complete the $D=2,4$ terms are taken from~\cite{art:Braaten}.


\section{RESULTS OF QCD FITS TO $\tau$ DATA} 
\begin{table*}[htb]
  \begin{center}
    \caption{\em Comparison of the QCD fit results (CIPT) to the
      moments of vector (axial-vector) current, the simultaneous fit
      of all moments for both currents, and the moments for the sum of
      both currents~\cite{art:SvenPhD}.  The quoted errors contain
      statistical errors only.}
    \begin{tabular}[t]{cllllllll}
      \hline
         & \multicolumn{2}{c}{V} & \multicolumn{2}{c}{A} & 
         \multicolumn{2}{c}{V and A } & \multicolumn{2}{c}{V + A }\\
         observable & \phantom{+}value & \phantom{+}error & 
         \phantom{+}value & \phantom{+}error & \phantom{+}value & 
         \phantom{+}error & \phantom{+}value & \phantom{+}error \\
      \hline               
   $\alphas(m_\tau^2)$   & 
   $\phantom{+}0.341$          & 
   $\pm 0.017$           &
   $\phantom{+}0.357$          & 
   $\pm 0.019$           &
   $\phantom{+}0.347$          & 
   $\pm 0.012$           &
   $\phantom{+}\MEANASVLMT$     &
   $\pm\MEANASXMMT$ \\
   $\langle\frac{\alphas}{\pi}\,GG\rangle/{\rm GeV^4}$ & 
   $\phantom{+}0.002$          & 
   $\pm 0.010$           &
   $-0.011$              & 
   $\pm 0.020$           &
   $\phantom{+}\MEANcGGVL$     & 
   $\pm 0.008$           &
   $\CIPTGGVL$           &
   $\pm\CIPTGGEM$\\
   $\delta_{\rm V}^6$    &  
   $\phantom{+}0.0259$         & 
   $\pm 0.0041$          &
     \phantom{+}---            & 
     \phantom{+}---            &
   $\phantom{+}\MEANcVSVL$     & 
   $\pm 0.0034$          & 
     \phantom{+}---            & 
     \phantom{+}---            \\
   $\delta_{\rm V}^8$& 
   $-0.0078$             & 
   $\pm 0.0018$          &
     \phantom{+}---            & 
     \phantom{+}---            &
   $\MEANcVEVL$          & 
   $\pm 0.0013$          &
     \phantom{+}---            & 
     \phantom{+}---            \\
   $\delta_{\rm A}^6$    &  
     \phantom{+}---            & 
     \phantom{+}---            &
   $-0.0246$             & 
   $\pm 0.0086$          &
   $\MEANcASVL$          & 
   $\pm 0.0033$          &
     \phantom{+}---            & 
     \phantom{+}---            \\
   $\delta_{\rm A}^8$    & 
     \phantom{+}---            & 
     \phantom{+}---            &
   $\phantom{+}0.0067$         & 
   $\pm 0.0050$          &
   $\phantom{+}\MEANcAEVL$     & 
   $\pm 0.0019$          &
     \phantom{+}---            & 
     \phantom{+}---            \\
   $\delta_{\rm V+A}^6$  &  
     \phantom{+}---            & 
     \phantom{+}---            &
     \phantom{+}---            & 
     \phantom{+}---            &
     \phantom{+}---            & 
     \phantom{+}---            &
     $\phantom{+}\CIPTSSVL$    &
     $\pm\CIPTSSEM$      \\
   $\delta_{\rm V+A}^8$  &  
     \phantom{+}---            & 
     \phantom{+}---            &
     \phantom{+}---            & 
     \phantom{+}---            &
     \phantom{+}---            & 
     \phantom{+}---            &
     $\CIPTSEVL$         &
     $\pm\CIPTSEEM$      \\
      \hline                                                
  $\chi^2/{\rm d.o.f.}$ &
  $\phantom{+}0.07/1$         & 
                        &
  $\phantom{+}0.06/1$         & 
                        &
  $\phantom{+}\CIPTcCHSQ/4$   &
                        &
  $\phantom{+} \CIPTCSMT/1$   &
                        \\
      \hline                       
    \end{tabular}
    \label{tab:V,A,VaA,VpA} \end{center}
\end{table*}                                
The fit results of QCD parameters to the ten moments $R_{\tau,{\rm
    V/A}}^{kl}$ for $kl=00,10,11,12,13$ as reported
in~\cite{art:Sven,art:SvenPhD} are shown in
Tab.~\ref{tab:V,A,VaA,VpA}.  Only the CIPT fits are presented here,
since the focus is the stability of the QCD fits and the two other
approaches lead to similar results within the theoretical
uncertainties. The various fits demonstrate the stability of the
method and that the perturbative parameter $\alpha_{\rm s}$ can be
measured together with the non-perturbative parts even if most of the
non-perturbative parts do not cancel as it is the case in the first
three fits.

Note that the statistical error on $\alpha_{\rm s}$ in the last two
fits is different only because the last fit uses additional
information from the $\tau$ lifetime and the leptonic branching
ratios~\cite{art:Sven}.

\begin{figure}[htb]
\centering
\resizebox{.5\textwidth}{!}{
\includegraphics{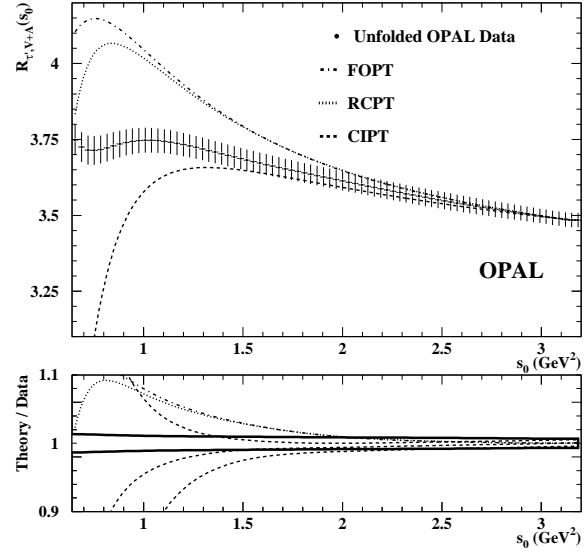}}
\caption{\em $R_{\tau'}^{V+A}(s_0)$ for a hypothetical $\tau'$ lepton
  versus the upper integration limit $s_0 = m_{\tau'}^2$.}
\label{fig:rtau}
\end{figure}
The extension to energy scales $s<m_\tau^2$ is demonstrated in
Fig.~\ref{fig:rtau}. Here the moment
$R_{\tau',V+A}^{00}(s_0=m_{\tau'}^2)$ is shown together with the fit
results to the three different theories FOPT, CIPT and RCPT for the
perturbative part plus non-perturbative parts.

The error on the theoretical curves (or alternatively their spread)
indicates that the OPE is applicable to values as low as $1.5\,{\rm
  GeV}^2$ which can be used to extract the running of the strong
coupling constant from $\tau$ data alone.

The three considered theories differ in the treatment of unknown
higher order terms which is partially taken into account as
theoretical uncertainty by means of the error on $K_4$. As a
conservative approach one could take the average of all three results
and their spread as additional theoretical uncertainty. From the
results in~\cite{art:Sven} one obtains therefore:
\begin{equation}\label{eq:asmtau}
  \alpha_{\rm s}(m_\tau^2) = 0.326 \pm 0.007_{\rm exp} \pm 0.022_{\rm theo},
\end{equation}
which translates to
\begin{equation}\label{eq:asmz}
  \alpha_{\rm s}(m_{\rm Z}^2) = 0.1194 \pm 0.0008_{\rm exp} \pm 0.0027_{\rm theo}
\end{equation}
at $m_{\rm Z}=91.188\,{\rm GeV}$.

\section{AVERAGING OF $\Pep\Pem$ DATA}
\begin{figure}[htb]
\centering
\resizebox{.5\textwidth}{!}{
\includegraphics{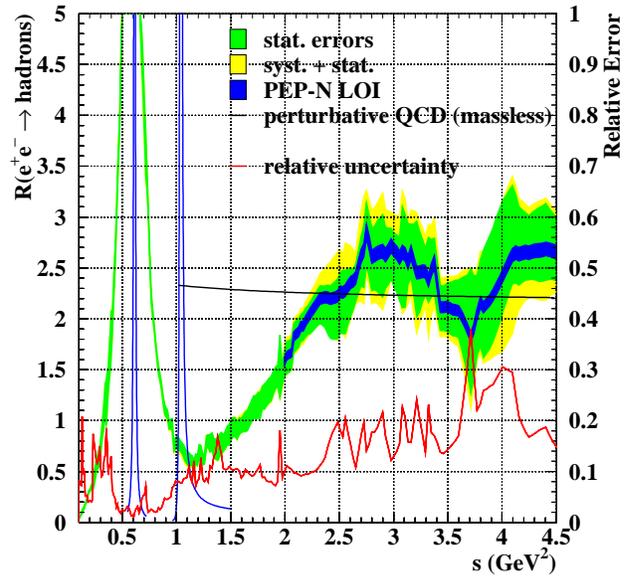}}
\caption{\em $R_{\Pep\Pem}(s)$ from exclusive channels. The dark gray
  band denotes statistical errors; the light gray band shows the sum
  of systematic and statistical errors. The dark solid line shows the
  relative uncertainty; the small dark error band beyond $s=2\,GeV^2$
  shows the projected uncertainty after 5 years of
  PEP-N~\cite{art:PEP-N}.  The narrow $\omega$ and $\phi$ resonances
  and the massless QCD prediction are also indicated by solid lines.}
\label{fig:repem}
\end{figure}

Exclusive $\Pep\Pem$ channels up to a CMS energy of $2.2\,{\rm
  GeV}$~\cite{Barkov:1985ac,Vasserman:1979hw,Vasserman:1981xq,Amendolia:1984di,Quenzer:1978qt,Bisello:1989hq,art:CMD2:2pi,Dolinsky:1991vq,Cosme:1976tf,Parrour:1976rt,Cosme:1979qe,Cordier:1980qg,Antonelli:1992jx,art:SND:2pipi0,Kurdadze:1986tc,Bisello:1991du,Esposito:1980by,Barkov:1988gp,Cordier:1982zn,art:CMD2:4pi,Antonelli:1988fw,Bisello:1981sh,Ivanov:1981wf,Ivanov:1982cr,Delcourt:1981eq,Mane:1981ep,Bisello:1988ez,Cordier:1982en,Mane:1982si}
are combined in this analysis to calculate the total hadronic cross
section.  The narrow $\omega$ and $\phi$ resonances are excluded from
the exclusive channels and added to the total cross section as
relativistic Breit-Wigner curves with $s$-dependent
widths~\cite{Eidelman:1995ny,art:PDG00}.  The weights of individual
channels and the treatment of systematic errors is taken
from~\cite{art:Davier} but the statistical errors and the final result
are obtained in a slightly different manner:
\begin{Enumerate}
\item a common equidistant binning in energy is chosen for all
  channels ($N=300$ bins from $E_{\rm min} = 0\,{\rm GeV}$ to $E_{\rm
    max} = \sqrt{5}\,{\rm GeV}$) and for every experiment and channel
  a histogram is filled with weighted averages of cross section
  measurements falling in the same bin:
  \begin{equation}
    \label{eq:trap}
    d_i = \frac{\displaystyle\sum_{\{j|N E_j/E_{\rm max}\in[i-1,i]\}}{
        \!\!\!\!\!\!\!\!\!\!\!\!\!\!\sigma_j/\Delta^2\sigma_j}}
    {\displaystyle\sum_{\{j|N E_j/E_{\rm max}\in[i-1,i]\}}{
        \!\!\!\!\!\!\!\!\!\!\!\!\!\!1/\Delta^2\sigma_j}},\quad i=1\dots N,
  \end{equation}
  where the squared errors $\Delta_i^2$ are given by the inverse of
  the denominator in~(\ref{eq:trap}).
\item gaps at $k=i\dots j$ between bins $i$ and $j$ of individual
  measurements are interpolated with the trapezoidal rule:
  \begin{equation}
    \label{eq:trap1}
    d_k = c_k d_i + (1-c_k) d_j,\quad c_k = 
    \frac{k-i}{j-i}.
  \end{equation}
\item the statistical errors are interpolated with the same procedure:
  \begin{equation}
    \label{eq:trap2}
    \Delta_k = c_k \Delta_i + (1-c_k) \Delta_j,
  \end{equation}
  and are therefore larger compared to Gaussian propagated errors by
  the factor:
  \begin{equation}
    \label{eq:trap3}
    r_k = \frac{c_k \Delta_i + 
      (1-c_k) \Delta_j}{c_k \Delta_i 
      \oplus (1-c_k) \Delta_j}.
  \end{equation}
\item the correlation matrix for the interpolated parts is given by:
  \begin{equation}
    \label{eq:trap4}
    \!\!\!\!V^{\rm stat}_{kl} = r_k r_l \left( c_k c_l \Delta_i^2 
      + (1-c_k)(1-c_l)\Delta_j^2\right).
  \end{equation}
\item the systematic errors are also interpolated according to the
  trapezoidal rule~(\ref{eq:trap2}) and all systematic errors are
  assumed to be $100\,\%$ correlated.
\item the final error matrix is given by the sum of both:
  \begin{equation}
    \label{eq:trap5}
    V_{kl} = V^{\rm stat}_{kl} + V^{\rm sys}_{kl}.
  \end{equation}
\item the averaged cross section over all experiments in bin $i$ for
  one exclusive channel is obtained from the weighted mean of all
  measurements and its error by usual error propagation. In case of
  inconsistent sets of data points for a particular bin its final
  error is scaled by the ratio $S = \sqrt{\chi^2/\chi^2_{68\,\%}}$, if
  $S>1$, where $\chi^2$ is the calculated $\chi^2$ for the individual
  experiments being consistent with the average and $\chi^2_{68\,\%}$
  is the $\chi^2$ value corresponding to a confidence level of
  $68\,\%$.
\end{Enumerate}
After this interpolation and averaging procedure the total hadronic
cross section is given by the sum of the exclusive channels as
in~\cite{art:Davier}.  The resulting values for $R_{\Pep\Pem}$ are
shown in Fig.~\ref{fig:repem}.


\section{RESULTS OF QCD FITS TO $\Pep\Pem$ DATA}
The five moments $R_{\Pep\Pem}^{kl}(4\,{\rm GeV}^2)$ for
$kl=20,30,31,32,33$ are given in Tab.~\ref{tab:repemkl}.  Their
correlations are given in Tab.~\ref{tab:repemklcorr}. The result of
the CIPT fit to these moments is given in Tab.~\ref{tab:repemfit}.
The correlations of the fit parameters are shown in
Tab.~\ref{tab:repemfitcorr}.
\begin{table}[htb]
  \begin{center}
    \caption{\em Moments $R_{\Pep\Pem}^{kl}(4\,GeV^2)$ for
      $kl=20,30,31,32,33$ from exclusive $\Pep\Pem$ data.  The errors
      include statistical and systematic uncertainties; projected
      errors after 5 years running of PEP-N and theoretical
      uncertainties are also given.}
    \begin{tabular}[t]{cllll}
      \hline
         $kl$ & $R_{\Pep\Pem}^{kl}$ & stat.+sys. & PEP-N  & theo.\\
      \hline               
        $20$  & $0.760$  & $0.014$  & $0.009$  & $0.008$ \\    
        $30$  & $0.569$  & $0.008$  & $0.007$  & $0.012$ \\
        $31$  & $0.1206$ & $0.0027$ & $0.0016$ & $0.0040$\\
        $32$  & $0.0355$ & $0.0015$ & $0.0006$ & $0.0000$\\
        $33$  & $0.0143$ & $0.0009$ & $0.0003$ & $0.0000$\\
      \hline
    \end{tabular}\label{tab:repemkl} 
  \end{center}
\end{table}

\begin{table}[htb]
  \begin{center}
    \caption{\em Correlations of the moments $R_{\Pep\Pem}^{kl}(4\,GeV^2)$ for
      $kl=20,30,31,32,33$ in percent. The values correspond to the
      quadratic sum of experimental and theoretical errors.}
    \begin{tabular}[t]{crrrr}
      \hline
         $kl$ & $30$     & $31$     & $32$     & $33$    \\
      \hline               
        $20$  & $+86.6$  & $ +3.8$  & $+74.5$  & $+70.4$ \\    
        $30$  &          & $-44.2$  & $+33.6$  & $+30.0$ \\
        $31$  &          &          & $+54.9$  & $+50.9$ \\
        $32$  &          &          &          & $+98.8$ \\
      \hline
    \end{tabular}\label{tab:repemklcorr} 
  \end{center}
\end{table}

\begin{table}[htb]
  \begin{center}
    \caption{\em Results from the QCD fits to the moments 
      $R_{\Pep\Pem}^{kl}(4\,GeV^2)$ for $kl=20,30,31,32,33$ from
      exclusive $\Pep\Pem$ data.  The errors include statistical and
      systematic uncertainties; projected errors after 5 years running
      of PEP-N and theoretical uncertainties are also given.}
    \begin{tabular}[t]{cllll}
      \hline
         obs. & \phantom{+}val. & err. & PEP-N  & theo.\\
      \hline               
        $\alpha_{\rm s}(4\,{\rm GeV}^2)$ & 
        $\phantom{+}0.286$  & $0.031$  & $0.027$  & $0.015$ \\ 
        $m_{\rm s}/{\rm GeV}$ & 
        $\phantom{+}0.220$  & $0.036$  & $0.026$  & $0.059$ \\
        $\langle{\cal O}_6\rangle/{\rm GeV}^6$ & 
        $-0.0041$ & $0.0007$ & $0.0005$ & $0.0002$\\
        $\langle{\cal O}_8\rangle/{\rm GeV}^8$ & 
        $\phantom{+}0.0043$ & $0.0004$ & $0.0002$ & $0.0002$\\
      \hline
        $\chi^2/{\rm d.o.f.}$ & $\phantom{+}0.04/1$ & & \\
      \hline
    \end{tabular}\label{tab:repemfit} 
  \end{center}
\end{table}

\begin{table}[htb]
  \begin{center}
    \caption{\em Correlations of the QCD fit parameters to the moments 
      $R_{\Pep\Pem}^{kl}(4\,GeV^2)$ for $kl=20,30,31,32,33$ in
      percent. The values correspond to the quadratic sum of
      experimental and theoretical errors.}
    \begin{tabular}[t]{crrr}
      \hline
         obs. & $m_{\rm s}$ & $\langle{\cal O}_6\rangle$ & 
        $\langle{\cal O}_8\rangle$ \\
      \hline               
        $\alpha_{\rm s}$            & $+77.2$  & $+71.4$  & $-42.0$ \\    
        $m_{\rm s}$                 &          & $+35.9$  & $-8.2$  \\
        $\langle{\cal O}_6\rangle$  &          &          & $-93.2$ \\
      \hline
    \end{tabular}\label{tab:repemfitcorr} 
  \end{center}
\end{table}
The value of the strong coupling at $4\,{\rm GeV}^2$ corresponds to
\begin{equation}\label{eq:asepemmz}
  \alpha_{\rm s}(m_{\rm Z}^2) = 0.117 \pm 0.005_{\rm exp} \pm 0.002_{\rm theo}
\end{equation}
at $m_{\rm Z}=91.188\,{\rm GeV}$ in good agreement with the $\tau$
result~\ref{eq:asmz} but with a larger experimental uncertainty.
\begin{figure}[htb]
\centering
\resizebox{.5\textwidth}{!}{
\includegraphics{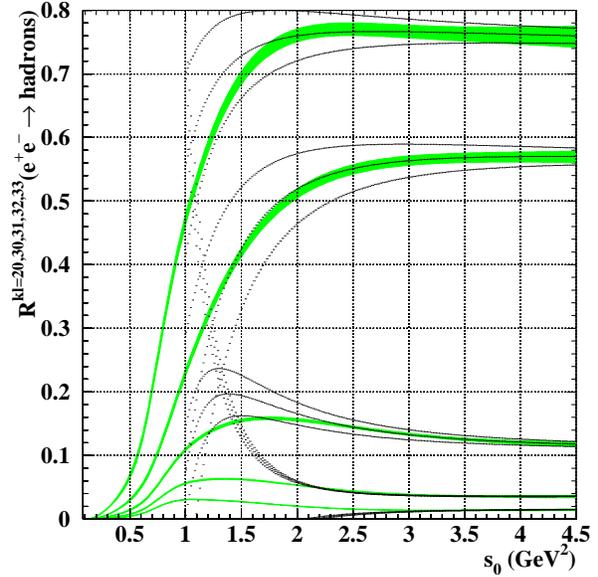}}
\caption{\em The moments $R_{\Pep\Pem}^{kl=20,30,31,32,33}(s_0)$
  versus the integration limit $s_0$. The shaded bands show the
  experimental moments with statistical and systematic uncertainties
  from top to bottom in the order $kl=20,30,31,32,33$. The dotted
  curves denote central values and $\pm 1\sigma$ ranges for the
  theoretical expectations (CIPT) using the fit values to the moments
  at $s_0=4\,GeV^2$ as input.}
\label{fig:epemfit}
\end{figure}
Figure~\ref{fig:epemfit} shows the moments of $R_{\Pep\Pem}$ as a
function of the upper integration limit together with theoretical
predictions using the fitted values at $s_0=4\,{\rm GeV}^2$ as input.
Reasonable agreement between the extrapolated and the measured moments
is observed for all five moments down to $s_0\approx2.5\,{\rm GeV}^2$.
The first three moments agree within errors with the extrapolated
results down to $s_0\approx1.5\,{\rm GeV}^2$.


\section{CONCLUSIONS}
The comparison of QCD fits to different currents (V, A, V and A, V+A)
in non-strange hadronic $\tau$ decays probes the stability of the OPE
at low energy scales. The good agreement of perturbative and
non-perturbative parameters among these fits demonstrates that QCD
fits can reliably be performed at these low energy scales.  The use of
spectral moments is therefore extended to $\Pep\Pem\rightarrow{\rm
  hadrons}$ data in order to fit QCD parameters at $4\,{\rm GeV}^2$.
While the extraction of the strong coupling $\alpha_{\rm s}$ is not
competitive in terms of experimental uncertainties with the
measurement from $\tau$ data it provides an important cross check and
probes an energy region not accessible with $\tau$ decays. Furthermore
the sensitivity to the mass of the strange quark can be used to
constrain this parameter. A value of $m_{\rm s}(1\,{\rm GeV}) =
(220\pm36\pm59)\,{\rm MeV}$ has been observed.  The experimental
uncertainty could be reduced by $30\,\%$ after 5 years of running with
PEP-N.

\section{ACKNOWLEDGMENTS}
I would like to thank the organizers Stan Brodsky and Rinaldo Baldini
for giving me the opportunity to present this work. I am also indebted
to Andreas H{\"o}cker who provided me with FORTRAN files of the $\Pep\Pem$
data.


%
%

\end{document}